# High-Performance Carbon Nanotube Transistors on SrTiO₃/Si Substrates

B. M. Kim[a),b)], T. Brintlinger[a)], E. Cobas[a)], Haimei Zheng[c)], and M. S. Fuhrer[a),d)]

*University of Maryland, College Park, Maryland 20742 USA*

Z. Yu, R. Droopad, J. Ramdani, and K. Eisenbeiser

*Physical Sciences Research Laboratories, Motorola Labs, Tempe, Arizona 85284*

Single-walled carbon nanotubes (SWNTs) have been grown via chemical vapor deposition on high-κ dielectric SrTiO₃/Si substrates, and high-performance semiconducting SWNT field-effect transistors have been fabricated using the thin SrTiO₃ as gate dielectric and Si as gate electrode. The transconductance per channel width is 8900 μS/μm. The high transconductance cannot be explained by the increased gate capacitance; it is proposed that the increased electric field at the nanotube-electrode interface due to the high-κ SrTiO₃ decreases or eliminates the nanotube-electrode Schottky barrier.

[a)]Department of Physics and Center for Superconductivity Research

[b)]Present address: Department of Mechanical Engineering and Applied Mechanics, University of Pennsylvania, Philadelphia, PA 19104

[c)]Department of Materials Science and Engineering,

[d)]Electronic mail: mfuhrer@physics.umd.edu

The high mobility, low defect structure, and intrinsic nanometer scale of semiconducting single-walled carbon nanotubes (SWNTs) has led to an intense research effort into the viability of SWNT field-effect transistors[1] (NT-FETs) as a replacement for, or complement to, future semiconductor devices. A number of researchers have attempted to improve the performance of NT-FETs by using thin, high-dielectric-constant (high-$\kappa$) dielectrics[2,3,4]. However, the presence of a Schottky barrier at the SWNT-metal interface has necessitated more unusual engineering solutions to obtain high transconductances, such as electrolytic gating[5] and local top-gating[4]. Ohmic contacts have also been achieved for the case of Pd[6] or Au[7] contacts and large-diameter nanotubes, but this solution may fail for the technologically-relevant small-diameter nanotubes that may be mass-produced[6].

We report here the integration of high-$\kappa$ SrTiO$_3$ on Si substrates (STO/Si) with NT-FETs. The chemical vapor deposition (CVD) of high quality SWNTs on STO/Si leaves the STO/Si intact with its desirable properties, and the NT-FETs demonstrate a high transconductance per width (8900 $\mu$S/$\mu$m), exceeding that reported for any other NT-FET[1-4,8]. We show that the increase in transconductance cannot be explained by an increase in the gate capacitance; the gate capacitance is largely limited by the quantum capacitance of the nanotube in our devices as well as other NT-FETs on high-$\kappa$ dielectrics[2-4]. We propose that the high transconductance of our devices is due instead to lowering or elimination of the Schottky barrier at the nanotube-metal interface by the high electric field at the dielectric-metal interface.

Our starting substrate consists of nominally 20-nm-thick epitaxial STO/Si ($\kappa \approx 175$). Details on the growth and characterization of SrTiO$_3$ on Si have been reported elsewhere[9,10]. SWNTs were grown by CVD, adapting from procedures by other researchers[11] for synthesis of SWNTs on SiO$_2$. Briefly, an alumina-supported Fe/Mo catalyst was patterned in islands on the substrate by electron-beam lithography[11]. CVD synthesis was carried out in a 1 in. diameter tube furnace for 11 min at 900 °C using a methane flow of ~1900 ml/min and a hydrogen co-flow of ~480 ml/min. To ensure the STO/Si remained intact after growth, we performed transmission electron microscopy (TEM) and electron diffraction. In Fig. 1(a) the STO and Si are readily identified in the TEM micrograph, along with an amorphous layer between. In Fig. 1(b), the electron diffraction pattern shows both Si and the STO reproducing results[12] for untreated STO/Si.

Evaporated Cr/Au electrodes (no annealing) were used to make source and drain contacts to the SWNTs. The Si served as a bottom gate electrode. The diameter $d$ of each nanotube, determined from atomic force microscopy (AFM), ranged from 1-10 nm. Presumably the sample comprises both single- and multi-walled nanotubes, but the smallest ($d < 1.3$ nm) nanotubes chosen for study of their device characteristics are almost certainly SWNTs[13].

In Figure 2, FESEM and AFM micrographs illustrate nanotube growth from a catalyst island. and the structure of the NT-FET. In Fig. 2(a), one sees a rough catalyst island, two Au/Cr leads, and several nanotubes, one which spans the leads. Figure 2(b) shows a similar nanotube device visualized by FESEM. In Fig. 2(c), the same area is imaged by AFM, more clearly resolving the nanotube. From the AFM topography, a diameter of ~1.0 nm is determined for this presumably single-walled nanotube. The gate length is 1.8 μm. The dielectric integrity of the $SrTiO_3$ was verified by measuring the current-voltage characteristic from the large-area (2.4 x $10^4$ μm$^2$) source and drain pads to the gate electrode for the device shown in Figs. 2(b) and (c). The gate leakage current does not exceed the noise level (~200 pA) for $V_s < \pm2$ V, and rises exponentially with $V_s$ for $V_s > \pm2$ V, to 2 x $10^{-4}$ A/cm$^2$ at 4 V, comparable to published leakage currents for similar substrates[9].

Figure 3 shows the drain current ($I_d$) vs. gate voltage ($V_{gs}$) of the device shown in Figs. 2(b) and (c). We numerically differentiate the data to calculate a transconductance $g_m = dI_d/dV_{gs}$. The $I_d(V_{gs})$ curves are sigmoidal, leading to a peak in $g_m$ as a function of $V_{gs}$. The inset of Fig. 3 shows this peak value of $g_m$ for each source voltage. The transconductance is approximately 8.9 μS at $V_{ds}$ = -800 mV. This value exceeds any reported for a globally-gated solid-state NT-FET[2,3,6,8] and is comparable to the value $g_m$ = 12 μS at $V_{ds}$ = -1200 mV for a locally-top-gated NT-FET with $ZrO_2$ dielectric[4]. In order to compare this value with other transistor technologies, we normalize by the device width[14] $d$. This leads to a transconductance per device width of $g_m/d$ = 8900 μS/μm. As seen in Table 1, $g_m/d$ exceeds all values in the literature for globally-gated NT-FETs, and also exceeds that for the locally-gated NT-FET[4] and electrolytically-gated NT-FET[5].

In a one-dimensional diffusive FET, the transconductance in the saturation region is given by $g_m \approx \mu c_g V_d/L$, where $c_g$ is the gate capacitance per length. Thus for a given material system (given $\mu$) the transconductance can be increased through increasing $c_g$ or $V_d$, or decreasing $L$. In practice, however, the product $\mu V_d/L$ is expected to have a maximum value $v_s$, the saturation carrier velocity, and the maximum transconductance is $g_{m,max} = c_g v_s$. Thus increasing $c_g$ becomes the goal for obtaining higher transconductance.

Table 1 compares the performance of the NT-FET on STO with other high performance NT-FETs. The transconductance of our NT-FET, 8.9 μS, is more than an order of magnitude greater than the values for NT-FETs on $Al_2O_3$ and $HfO_2$ dielectrics, 0.3 and 0.6 μS respectively[2,3]. However, these differences in transconductance observed in NT-FETs cannot be explained by increased gate capacitances within the diffusive FET model, as follows. The electrostatic gate capacitance per length may be approximated by $c_{g,el} = 2\pi\kappa\varepsilon_o/\ln(4t/d)$ where $t$ is the dielectric thickness, and $\kappa$ the dielectric constant (this formula somewhat overestimates the capacitance, due to the lack of dielectric above the nanotube). The total gate capacitance $c_g$ must take into account the quantum capacitance $c_q$ of the nanotube; $c_g = c_{g,el}c_q/(c_{g,el}+c_q)$, where $c_q \approx 4$ pF/cm[5]. Thus $c_g$ is dominated by the smaller of $c_{g,el}$ and $c_q$; as $c_{g,el}$ is increased, $c_g$ tends to the value 4 pF/cm. Table 1 gives values for $t$, $d$, $L$, $V_{sd}$, $c_{g,el}$ and $c_g$. For the high-dielectric-constant devices[2-4], $c_{g,el}$ is comparable to or exceeds $c_q$, and hence $c_g$ is on order $c_q$. Thus the observed ~15-30x variation in $g_m$ cannot be explained by a ~3-5x variation in $c_g$ in the standard, diffusive FET model. If we further consider that the transconductance has not reached its saturation value in our experiment or in Refs. 2 or 3, the greater $L$ and smaller $V_d$ of our device compared to those in Refs. 2 or 3 should result in an even lower transconductance.

The failure of the diffusive FET model in NT-FETs is not surprising; as other researchers have pointed out, the transconductance of NT-FETs is often controlled by Schottky barriers at the nanotube-metal interface[15]. This readily explains the much higher transconductances observed in Refs. 4 and 6, where the effects of the Schottky barriers at the nanotube-metal contact were circumvented through local-top-gating or Ohmic contacts to the nanotube, respectively. A simple electrostatic model predicts that the transconductance for Schottky-barrier NT-FETs scales as the inverse square-root of the dielectric thickness, $g_m \sim t^{-1/2}$, and surprisingly is independent of the dielectric constant[16,17]. This result is also inadequate to explain the differences in transconductances[18] in Tab. 1; our device has $t$ comparable to or larger than refs. 2 and 3.

One possible explanation for the discrepancy is that we have made Ohmic contact to the nanotube. This is in contrast to Schottky-barrier-FET behavior observed for similarly small $d$ nanotubes in Ref. 6, and as-deposited (not annealed) Cr/Au contacts in Ref. 7. We note also that the subthreshold swing in our devices is ~400 mV/decade at room temperature (not shown), much larger than the 150-170 mV/decade observed for Ohmically-contacted NT-FETs in Ref. 6, though this could also result from a larger interface trap density in STO.

Another possibility is that vertical scaling has a more pronounced effect on the Schottky barriers. The model of Ref. 16 may be inadequate for two reasons. First, Ref. 16 ignores the charge in the nanotube channel. A self-consistent treatment of charge in the on-state does show increases in device on-current for increased dielectric constants[19]. Second, Ref. 16 treats the nanotube as infinitely thin. We expect that the details of the electric field at the contacts will be substantially modified when the effective thickness of the dielectric $t' = t/\kappa$ becomes significantly less than the nanotube diameter $d$ (the dielectric constant of the nanotube is unity[20]). In our devices, $t'/d \approx 0.1$, significantly less than the values of $t'/d \approx 0.2$-$0.6$ and $1$-$2$ in refs. 2 and 3 respectively. (For typical devices fabricated on thick $SiO_2$, $t'/d \approx 100$.) When $t'/d << 1$, the potential drop *across* the nanotube diameter for the portion of nanotube underneath the electrode becomes a large fraction of the applied gate voltage. When the voltage drop across the radius of the nanotube is equal to the Schottky barrier height, population of the valence band with carriers should become energetically favorable, allowing Ohmic contact with the channel. Stated another way, at moderate gate voltages the shift in electrostatic potential of the nanotube relative to the metal electrode can be greater than the Schottky barrier height, eliminating the barrier. In our devices this would occur at an applied gate voltage of a few hundred mV from threshold. This model offers an alternate explanation for the observation of high transconductances (even in small diameter nanotubes) in FETs with an electrolyte dielectric[5] ($t'/d \approx 0.01$).

Acknowledgements:

This research was supported by ARDA and the Office of Naval Research through grant N000140110995, the Director of Central Intelligence Postdoctoral Research Fellowship Program, and the National Science Foundation through grant DMR-0102950. The authors are grateful for helpful conversations with Frank D. Gac and Ramamoorthy Ramesh.

Table I: Device parameters for high-transconductance nanotube field-effect transistors in this work and other works. The columns display the dielectric material and dielectric constant $\kappa$, dielectric thickness $t$, nanotube diameter $d$, gate length $L$, electrostatic gate capacitance $c_{g,el}$, total gate capacitance $c_g$, transconductance $g_m$, source-drain bias $V_{sd}$ and transconductance per width. The symbol ‡ denotes electrolytic gating, * local top gating, and † Ohmic contacts.

| Author | Dielectric ($\kappa$) | $t$ (nm) | $d$ (nm) | $L$ ($\mu$m) | $c_{g,el}$ (pF/cm) | $c_g$ (pF/cm) | $g_m$ ($\mu$S) | $V_{ds}$ (V) | $g_m/d$ ($\mu$S/$\mu$m) |
|---|---|---|---|---|---|---|---|---|---|
| Bachtold (Ref. 2) | $Al_2O_3$ (5) | 2-5 | 1 | 0.2 | 0.9-1.3 | 0.7-1.0 | 0.3 | -1.3 | 300 |
| Appenzeller (Ref. 3) | $HfO_2$ (11) | 20 | 1-2 | 0.3 | 1.6 | 1.1 | 0.6 | -1.5 | 300-600 |
| This work | $SrTiO_3$ (175) | 20 | 1.0 | 1.8 | 22 | 3.4 | 8.9 | -0.8 | 8900 |
| Rosenblatt‡ (Ref. 5) | Electrolyte (80) | ~1 | 3 | 1.4 | 70 | 3.8 | 20 | -0.8 | 6700 |
| Javey* (Ref. 4) | $ZrO_2$ (25) | 8 | 2 | 2 | 5.5 | 2.3 | 12 | -1.2 | 6000 |
| Javey† (Ref. 6) | $SiO_2$ (3.9) | 500 | 3.3 | 0.3 | 0.34 | 0.31 | 5 | -0.6 | 1540 |

Figure Captions

Fig. 1. Transmission electron (a) micrograph and (b) diffraction pattern. In (a), the crystalline $SrTiO_3$, amorphous interface layer, and crystalline Si substrate are resolved in profile. In (b), two sets of diffraction spots reveal the presence of $SrTiO_3$ and Si.

Fig. 2. Images of carbon nanotubes grown on $SrTiO_3$ (STO) substrates by chemical vapor deposition. (a) showing patterned catalyst (left) on STO, as well as several nanotubes extending from the catalyst island. One nanotube has been contacted by two Cr/Au electrodes. (b) Field-emission scanning electron micrograph of a semiconducting nanotube on STO bridging two Cr/Au contacts with 1.8 um separation. (c) AFM image of the nanotube in (b).

Fig. 3. Drain current ($-I_d$) as a function of gate voltage ($V_{gs}$) at drain voltages ($V_{ds}$) of -100 mV to -800 mV in 100 mV steps, at room temperature. Inset shows the maximum transconductance $dI_d/dV_{gs}$ calculated from these data.

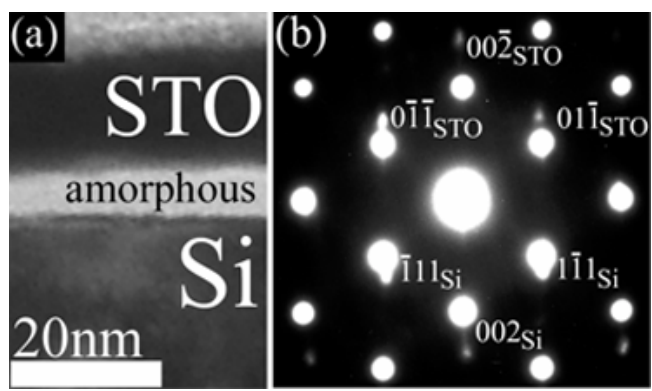

Figure 1

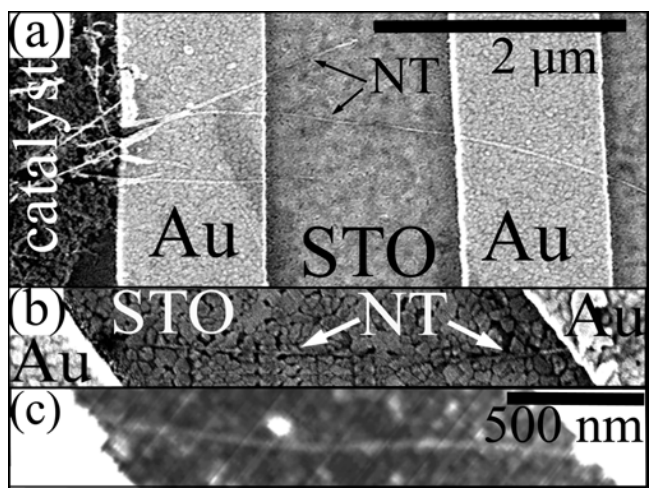

Figure 2

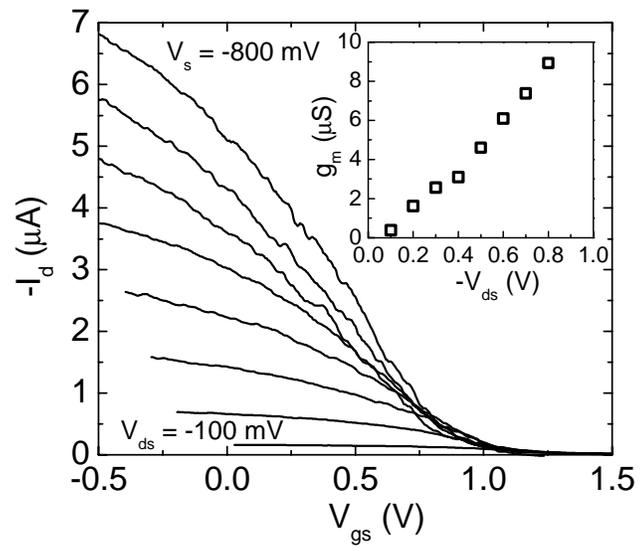

Figure 3